\documentstyle[aps]{revtex}

\begin{document}
\title{What Casimir Energy can suggest about Space Time Foam?}
\author{Remo Garattini\thanks{%
Talk given at the {\it Fifth Workshop on Quantum Field Theory under the
Influence of External Conditions, }Leipzig, September 10-14, 2001}}
\address{Facolt\`{a} di Ingegneria, Universit\`{a} degli Studi di Bergamo,\\
Viale Marconi, 5, 24044 Dalmine (Bergamo) Italy.\\
E-mail:\\
Garattini@mi.infn.it}
\maketitle

\begin{abstract}
In the context of a model of space-time foam, made by $N$ wormholes we
discuss the possibility of having a foam formed by different configurations.
An equivalence between Schwarzschild and Schwarzschild-Anti-de Sitter
wormholes in terms of Casimir energy is shown. An argument to discriminate
which configuration could represent a foamy vacuum coming from Schwarzschild
black hole transition frequencies is used. The case of a positive
cosmological constant is also discussed. Finally, a discussion involving
charged wormholes leads to the conclusion that they cannot be used to
represent a ground state of the foamy type.
\end{abstract}

\vspace{3cm}\noindent{\bf Problem}: How calculate the zero point energy
(Z.P.E.) in pure gravity and what is the Casimir effect in such a
configuration? The formulation of the Casimir effect in general is
synthetized by the Eq.
\begin{equation}
E_{Casimir}\left[ \partial {\cal M}\right] =E_{0}\left[ \partial {\cal M}%
\right] -E_{0}\left[ 0\right] ,
\end{equation}%
where $E_{0}$ is the Z.P.E. and $\partial {\cal M}$ is a boundary. It is
immediate to see that the Casimir energy involves a vacuum subtraction
procedure and since this one is related to Z.P.E., we can extract
information on the ground state. Space-time foam is a possible candidate for
a ground state of quantum gravity. It was J. A. Wheeler who first
conjectured that spacetime could be subjected to topology fluctuation at the
Planck scale\cite{Wheeler}. This means that spacetime undergoes a deep and
rapid transformation in its structure. This changing spacetime is best known
as space-time foam. However, in which way it is possible to construct such a
fluctuating structure. One possibility comes by the computation of the
following quantity
\[
E\left( wormhole\right) =E\left( no-wormhole\right)
\]%
\begin{equation}
+\Delta E_{no-wormhole}^{wormhole}{}_{|classical}+\Delta
E_{no-wormhole}^{wormhole}{}_{|1-loop},  \label{i0}
\end{equation}%
representing the total energy computed to one-loop in a wormhole background.
$E\left( no-wormhole\right) $ is the reference space energy which, in the
case of the Schwarzschild and RN wormhole, is flat space. $\Delta
E_{no-wormhole}^{wormhole}{}_{|classical}$ is the classical energy
difference between the wormhole and no-wormhole configuration stored in the
boundaries and finally $\Delta E_{no-wormhole}^{wormhole}{}_{|1-loop}$ is
the quantum correction to the classical term. One reason leading to Eq.$%
\left( \ref{i0}\right) $ is in vacuum Einstein equations
\begin{equation}
R_{\mu \nu }-\frac{1}{2}g_{\mu \nu }R=0.  \label{i1}
\end{equation}%
The only spherically symmetric metrics solutions of Eq.$\left( \ref{i1}%
\right) $ are the flat and Schwarzschild metric, respectively. The
Schwarzschild metric can be thought as a wormhole with topology $S^{2}\times
R^{1}$ which asymptotically tend to the flat metric. Therefore, in this
context, it is natural the comparison between the Schwarzschild wormhole and
the flat space Z.P.E.. The inclusion of a negative and positive cosmological
constant is straightforward. Thus the main reason leading to Eq.$\left( \ref%
{i0}\right) $ is that the wormhole is a measure of the strong curvature of
the gravitational field and the related Casimir energy is a measure of the
vacuum energy. In a series of papers, we have used such an idea to
concretely realize a model of space-time foam composed by a collection of $%
N_{w}$ Schwarzschild wormholes \cite{Remo,Remo1,Remo2,Remo3,Remo4,Remo5}. A
consequence of this model is that if we compute the area of the
Schwarzschild black hole event horizon on a ``{\it foam}'' state $\left|
\Psi _{F}\right\rangle $ we get%
\begin{equation}
M=\frac{\sqrt{N}}{2l_{p}}\sqrt{\frac{\ln 2}{\pi }},  \label{i2}
\end{equation}%
namely the Schwarzschild black hole mass $M$ has been {\it quantized} in
terms of $l_{p}$\cite{Ahluwalia,Hod,Makela,VazWit,Mazur,Kastrup,JGB}. This
implies also that the level spacing of the transition frequencies is
\begin{equation}
\omega _{0}=\Delta M=\left( 8\pi Ml_{p}^{2}\right) ^{-1}\ln 2.  \label{i3}
\end{equation}%
Note that for the Schwarzschild wormholes, Eq.$\left( \ref{i0}\right) $
gives at its minimum
\begin{equation}
\Delta E_{N_{w}}=-N_{w}\frac{V}{64\pi ^{2}}\frac{\Lambda ^{4}}{e},
\label{i4}
\end{equation}%
obtained by computing the total Casimir energy
\[
N_{w}\Delta E=N_{w}\left( E^{wormhole}-E^{No-wormhole}\right)
\]%
\begin{equation}
=\frac{N_{w}}{2\pi }\int_{0}^{+\infty }dp\int_{0}^{+\infty }dl\left(
2l+1\right) \left[ \left( \frac{d\delta _{l}^{+}\left( p\right) }{dp}+\frac{%
d\delta _{l}^{-}\left( p\right) }{dp}\right) \right] p.  \label{i5}
\end{equation}%
The phase shift is defined as $\left( r\equiv r\left( x\right) \right) $%
\begin{equation}
\delta _{l}^{\pm }\left( p\right) =\lim_{R\rightarrow +\infty }\left[
\int_{r_{h}}^{x\left( R\right) }dx\sqrt{p^{2}-\frac{l\left( l+1\right) }{%
r^{2}}-\tilde{V}^{\mp }\left( r\right) }-\int_{r_{h}}^{x\left( R\right) }dx%
\sqrt{p^{2}-\frac{l\left( l+1\right) }{r^{2}}}\right] ,  \label{i6}
\end{equation}%
where $x$ is the proper distance from the throat and $\tilde{V}^{\mp }\left(
r\right) $ is the curvature potential due to the wormhole. In Eq.$\left( \ref%
{i4}\right) $ a cut-off $\Lambda $ has been introduced to keep under control
the U.V. divergence and $V$ is a ``local'' volume defined by
\begin{equation}
V=4\pi \int_{x\left( r_{h}\right) }^{x\left( r_{0}\right) }dxr^{2}.
\label{i7}
\end{equation}%
$r_{h}$ is the throat location (horizon) and $r_{0}>r_{h}$ with $r_{0}\neq
\alpha r_{h}$ and $\alpha $ is a constant. It is interesting to see that Eq.$%
\left( \ref{i4}\right) $ appears even for the Anti-de Sitter (AdS) case,
described by the Schwarzschild-Anti-de Sitter (S-AdS) wormholes\cite{RCQG1}.
This means that our model of foam can be represented either by Schwarzschild
wormholes or by S-AdS wormholes. If we compute the Schwarzschild black hole
level spacing of transition frequencies in terms of S-AdS wormholes, we
obtain
\begin{equation}
\omega _{0}=\Delta M_{S}^{AdS}=\frac{9}{16}\left( 8\pi Ml_{p}^{2}\right)
^{-1}\ln 2
\end{equation}%
which is smaller with respect to $\omega _{0}$ of Eq.$\left( \ref{i3}\right)
$. Thus we can use the difference in the spectrum to select the correct
configuration of the foam constituents. Of course, nothing prevents to
consider even a positive cosmological constant. The associated wormhole
metric is described by the Schwarzschild-de Sitter metric which contains two
throats: the wormhole throat and the cosmological throat. This last one
contributes to the Z.P.E. leading to a value of\cite{RCQG2}
\begin{equation}
\Delta E_{N_{w}}=-N_{w}\frac{V}{32\pi ^{2}}\frac{\Lambda ^{4}}{e}.
\end{equation}%
Nevertheless the case of the positive cosmological constant cannot be
directly compared with the configuration spaces leading to Eq.$\left( \ref%
{i4}\right) $, because the effect of quantum fluctuation is that of inducing
a cosmological constant, which in the case of the Schwarzschild-de Sitter
wormhole is included from the beginning. The situation seems more closely
related to a sequence decay mechanism of the type\newpage

\[
Flat\ Space
\]%
\[
\Downarrow
\]%
\[
Schwarzschild\ Spacetime\ Foam
\]%
\[
\Downarrow
\]%
\[
Schwarzschild-de\ Sitter\ Spacetime\ Foam.
\]%
Finally we wish to mention the case with a charge without a cosmological
constant. In this situation we have the electric field contribution
(magnetic field, respectively) to the Z.P.E.. Due to the presence of this
non-gravitational field to Z.P.E., one finds that the Z.P.E. for charged
wormholes is always higher than the neutral ones\cite{Remo6}. A similar
computation in presence of a cosmological constant has not been done.
However, the effect of the cosmological constant on Z.P.E. is to shift the
vacuum position which is subsequently subtracted by the reference space.
This is the main reason that leads to a common result for the AdS and the
Schwarzschild wormholes. An important remark corroborating the foam vacuum
is that a shift of the Z.P.E. with respect to the expected vacuum should not
be there. The fact that one discovers a deviation from the expected result,
even in absence of a renormalization process, is a clear signal of a bad
vacuum choice.

\section{Acknowledgments}

I would like to thank Prof. M. Bordag who has given to me the opportunity of
participating to the Conference and Prof. A. Perdichizzi for a partial
finantial support.

\eject

\end{document}